\documentstyle[aps,12pt]{revtex}

\begin{document}
\title{Transmission time of wave packets through tunneling barriers}
\author{Yu. E. Lozovik\thanks{%
E-Mail: lozovik@isan.troitsk.ru} and A. V. Filinov\thanks{%
E-Mail: alex@vovan.msk.ru}}
\address{Institute of Spectroscopy, Russian Academy of Sciences\\
142092 Troitsk, Moscow District, Russia}
\date{Zh. \'{E}ksp. Teor. Fiz. {\bf 115}, 1872--1889 (May 1999}
\maketitle

\begin{abstract}
The transmission of wave packets through tunneling barriers is studied in
detail by the method of quantum molecular dynamics. The distribution
function of the times describing the arrival of a tunneling packet in front
of and behind a barrier and the momentum distribution function of the packet
are calculated. The behavior of the average coordinate of a packet, the
average momentum, and their variances is investigated. It is found that
under the barrier a part of the packet is reflected and a Gaussian barrier
increases the average momentum of the transmitted packet and its variance in
momentum space.

\vspace{0.25in}
\end{abstract}

\section{Introduction}

The study of tunneling in nanostructures has assumed an important role in
the last few years in connection with advances in nanoelectronics. The
problem of tunneling of wave packets through a potential barrier arises in
many cases, for example, in the study of the action of femtosecond light
pulses on coupled wells. This problem is also important because of possible
applications of scanning tunneling microscopes irradiated with femtosecond
pulses for simultaneously high-spatial and high-temporal resolution study of
nanostructures.\cite{r1} Other interesting questions are the tunneling time
in the ionization of a hydrogen atom by ultrashort laser pulses and the
tunneling time for tunneling induced by the action of a laser pulse on
low-lying nuclear energy levels. In the present paper we investigate the no
less interesting question of the time duration of tunneling in nanostructures.
The tunneling time is of practical interest in this case because it permits
to estimate the response time of semiconductor components. In this connection
we shall study the following problem: Let a laser pulse produce a wave
packet of an excited electron near a tunneling barrier. The question is:
When will the tunneling portion of the packet appear behind the barrier? The
arrival of the wave packet can be detected by studying local variations of
the optical properties using ultrashort probe pulses.

It is interesting that a number of effects which are absent in the
stationary case are observed when a packet passes through a tunneling
barrier. The tunneling time of a packet is determined in general not by the
reciprocal of the probability of stationary tunneling, but rather it is
related with quite complicated processes --- a change in the shape and
behavior of the packet inside the barrier. Moreover, the transmission time
through a barrier depends on the measured quantities, i.e. on
the type of experiment.

The investigation of the question of the presence time of a tunneling
particle under a barrier started quite a long time ago,\cite{r2,r3,r4} and
many theoretical and experimental methods for measuring the tunneling time
have been proposed. For example, there exist approaches where the peak of
the packet or the average coordinate (the ``centroid'') is chosen as the
observed quantity while the reflection or transmission time is determined by
their evolution. However, it has been shown in Refs. \onlinecite{r5} and \onlinecite{r6} and subsequent
works that a wave-packet peak incident on a potential barrier does not pass
into the peak of the transmitted wave. In Ref. \onlinecite{r7} it was shown that
on account of dispersion of the wave packet in momentum space the high-energy
components reached the barrier before the other components. Since the
tunneling probability increases with energy, these components made the main
contribution to the transmitted part of the packet. The initial parameters
could be chosen in a manner so that the transmitted part of the packet left
the barrier long before the main peak, chosen as the observed quantity,
appears. This example demonstrates the breakdown of the causality principle,
which is the basis of this method, and therefore it limits the applicability
of the method. Moreover, it is difficult to conceive of an experimental
method for measuring the arrival time of a packet according to its peak or
``centroid.''

There also exists a class of approaches that employ an ensemble of dynamical
trajectories to determine the tunneling time. These dynamical trajectories
arise as a necessary apparatus of the description in the Feynman and Bohm
interpretations of quantum mechanics. When Feynman trajectories were used,%
\cite{r8} the transmission time through a barrier was determined as a path
integral over all possible trajectories that start from a prescribed point
to the left of the barrier and arrive at a certain time at a point located
to the right of the barrier. The integrated function in the path integral
contained the product of a classical presence time of a trajectory inside
the barrier by a weighting factor $\exp \{iS(x(t^{\prime }))/\hbar ]$, where
$S(x(t^{\prime }))$ is the action related with the trajectory $x(t^{\prime })
$ under consideration. The calculated times possess real and imaginary parts
because of the multiplication by a complex weighting factor, and the
question of how these times should be associated to the physically
observable quantities, which are always real, arose. To explain the complex
times, which also arise in other methods (for example, in the method of
``physical clocks'' (see below)), Sokolovski and Connor\cite{r9} examined
so-called direct and indirect measurements. In indirect measurements as in
the case of Feynman trajectories method the quantities can be complex.

Approaches employing physical clocks have found quite wide application.
Physical clocks are various additional degrees of freedom in the system that
make it possible to determine the presence time of a particle in a given
region. Three types of clocks have been investigated in theoretical works.
Baz' and Rybachenko\cite{r10,r11} used spin precession in a weak uniform
magnetic field applied inside a barrier. At first spin precession in a single plane
was considered. Then Buttiker and Landauer\cite{r12} extended the analysis
to three dimensions. During the tunneling a spin of a particle acquires a component
along the direction of motion and along the magnetic field. It is obvious
that the intensities of the detected components with spin polarization in
these two directions will be proportional to the presence time of the
particle in the region with the magnetic field, i.e. in the region of the
barrier. It turned out that for a square barrier the tunneling times
determined in this manner are identical to the real and imaginary parts of
the complex tunneling time introduced via Feynman path integrals.\cite{r8}
The extension of this method to the case of an arbitrary potential barrier
was made in Ref. \onlinecite{r13}. Buttiker and Landauer\cite{r14} considered as physical
clocks an oscillating barrier in which the amplitude of the oscillations of
the temporal component was much smaller than the barrier height. At low
frequencies particles see an effective static barrier, since the
transmission time through the barrier is much shorter than the period of the
oscillations of the temporal component of the barrier. As the frequency
increases, the delayed particles or wave-packet components see a slightly
modified potential barrier. Finally, for some frequencies one or several
periods of the oscillations influence the tunneling particles. The frequency
at which a substantial difference from the adiabatic case corresponding to a
stationary barrier appears will determine the reciprocal of the
interaction time with the barrier or the transmission time through the
barrier. Martin and Landauer\cite{r15} chose as physical clocks the
oscillating amplitude of the incident wave. For this, a wave function
consisting of a superposition of two plane waves with different energies was
chosen to the left of the barrier. It is obvious that in this case the wave
function to the right of the barrier will also be a superposition of the
tunneled parts of the plane waves, which, however, possess a different
transmission amplitude, since the amplitude depends on the energy. The
transmitted wave function will reproduce the incident wave function if the
amplitudes of the tunneled plane waves differ very little; this corresponds
to the adiabatic case. The energy difference between the initial plane waves
for which the wave function behind the barrier does not reproduce the incident
wave function makes it possible to determine the transmission time through a
potential barrier. The main advantage of this method is that it is
applicable for all types of potentials, but it employs two values of the
energy, so that it is not clear to which energy the tunneling time obtained
should be ascribed.

Do all clocks give the same measurement result? Of course, no. However, in
many cases these results are close to one another or identical.\cite
{r13,r16,r17,r18} The main advantage of the approaches using physical clocks
is that they strive to determine the tunneling time in terms of possible
measurements in physical expeirments.

The search for ``time operators'' and the study of their properties is no
less popular.\cite{r19,r20,r21,r22,r23,r24} As first noted by Pauli,\cite
{r25} the main difficulty is that a measurable hermitian time operator
for a system Hamiltonian with a bounded spectrum does not exist. Various
attempts have been made to construct operators that would describe the
necessary properties of physical times. In order that the constructed
operator satisfy the correspondence principle, relations from classical
mechanics were taken as the basis for the operator construction. However, it
is well known that the construction of an operator expression corresponding
to a classical quantity is not unique, and its relation with the measurement
process requires additional analysis.

In the present work the tunneling time was determined as the difference of
the average ``arrival'' and ``presence'' times of a wave packet (see Sec.
3) before and after the barrier. The method of quantum molecular dynamics
was used to calculate these times and to investigate the dynamics of a
tunneling wave packet.\cite{r26,r27,r28}

It is well known that molecular dynamics investigates the properties of
classical systems in phase space. Therefore it is natural to extend
this method to quantum systems in phase space. The evolution of a system in
phase space can be described, for example, on the basis of the Wigner
formalism of quantum mechanics by the Wigner--Liouville equation. To solve
the Wigner--Liouville equation written in integral form it is convenient to
rewrite the equation in the form of an iterational series. Each term of this
series can be treated as a weighted contribution of a trajectory consisting
of segments of classical trajectories separated by finite disturbances of
the momentum. In what follows we shall call such a trajectory a quantum
trajectory. The statistical ensemble of quantum trajectories makes it
possible to calculate the sum of all terms in the series. The Monte Carlo
method is used to take account of only the trajectories making the main
contribution. In the classical limit the quantum trajectories turn into
classical trajectories, and the method of generalized molecular dynamics
becomes identical to usual molecular dynamics. The principles of the
method are presented in Sec. 2. The expressions for calculation of the
distributions of the arrival and presence times of a wave packet are
presented in Sec. 3 on the basis of the Wigner formalism of quantum
mechanics. The simulation results are discussed in Sec. 4. The
one-dimensional case is considered in this paper, but the method employed
makes it possible to perform similar calculations for multidimensional and
multiparticle systems, where it has serious advantages from the standpoint
of computer time over, for example, the solution of the nonstationary
Schr\"{o}dinger equation.

\section{Computational method}

To calculate the quantum-mechanical average of a quantity $A$ for a
nonstationary state $\left| \psi \right\rangle $ in the Wigner formulation
of quantum mechanics it is necessary to calculate an integral in phase space%
\cite{r29}

\begin{equation}
A\left( t\right) =\left\langle \psi \left| \hat{A}\left( t\right) \right|
\psi \right\rangle =\int \!\!\!\int dqdpA\left( q,p\right) W\left(
q,p,t\right) ,
\end{equation}
where, by definition, the Weyl symbol $A(q,p)$ is introduced for the
operator $\hat{A}$ and $W(q,p,t)$ is the Wigner function, which is the
Fourier transform of the off-diagonal density-matrix element:

\begin{eqnarray}
A\left( q,p\right)  &=&\int d\xi \exp \left( \frac{ip\xi }\hbar \right)
\left\langle q+\frac \xi 2\left| \hat{A}\right| q-\frac \xi 2\right\rangle ,
\\
W\left( q,p,t\right)  &=&\frac 1{2\pi \hbar }\int d\xi \exp \left( -\frac{%
ip\xi }\hbar \right) \psi ^{*}\left( q-\frac \xi 2,t\right) \psi \left( q+%
\frac \xi 2,t\right) .
\end{eqnarray}
Differentiating the distribution function with respect to time, substituting
it for the time derivative of the function $\psi $ on the right-hand side of
the Schr\"{o}dinger equation, and integrating by parts, we obtain the
Wigner--Liouville integrodifferential equation\cite{r30}

\begin{equation}
\frac{\partial W}{\partial t}+\frac pm\,\frac{\partial W}{\partial q}%
+F\left( q\right) \frac{\partial W}{\partial p}=\int\limits_{-\infty
}^\infty dsW\left( p-s,q,t\right) \omega \left( s,q\right) .
\end{equation}
In this equation

\begin{equation}
\omega \left( s,q\right) =\frac 2{\pi \hbar ^2}\int dq^{\prime }V\left(
q-q^{\prime }\right) \sin \left( \frac{2sq^{\prime }}\hbar \right) +F\left(
q\right) \frac{d\delta \left( s\right) }{ds}
\end{equation}
takes account of the nonlocal contribution of the potential, and $%
F(q)=-\partial V(q)/\partial q$ is a classical force. In the classical
limit, $\hbar \rightarrow 0$, Eq. (4) becomes the classical Liouville
equation

\begin{equation}
\frac{\partial W}{\partial t}+\frac pm\,\frac{\partial W}{\partial q}=-F
\left( q\right) \frac{\partial W}{\partial p}.
\end{equation}

The equation (4) can be written in an integral form. For this, one
introduces the dynamical trajectories $\{\bar{q}_\tau (\tau ;p,q,t),\bar{p}%
_\tau (\tau ;p,q,t)\},$ $\tau \in [0,t]$, starting from the point $(p,q)$ at
time $\tau =t$:

\begin{eqnarray}
d\bar p/d\tau &=&F(\bar p(\tau )),\ \bar
p_t(t;p,q,t)=p \nonumber
\\ d\bar q/d\tau &=&\bar q(\tau )/m,\ \bar
q_t(t;p,q,t)=q
\label{s8}
\end{eqnarray}

An integral equation is obtained by substituting the right-hand sides of
these equations into the Wigner--Liouville equation, whose left-hand side
becomes a total differential, and integrating over time one have

\begin{equation}
W\left( p,q,t\right) =W^0(\bar{p}_0,\bar{q}_0)+\int\limits_0^td\tau
\int\limits_{-\infty }^\infty dsW\left( \bar{p}_\tau -s,\bar{q}_\tau ,\tau
\right) \omega \left( s,\bar{q}_\tau \right) .
\end{equation}
Here $W^0(\bar{p}_0,\bar{q}_0)=W(p,q,0)\,\,$is the Wigner distribution
function at zero time. The solution of Eq. (8) can be represented as an
iterational series. For this we introduce the notation $\tilde{W}^{\tau _1}$
for the distribution function, which evolves classically in the interval $%
[0,\tau _1]$, and the integral operator $K_{\tau _i}^{\tau _{i+1}}$
describing the evolution between the times $\tau _i$ and $\tau _{i+1}$. Now
Eq. (8) can be represented in the form

\begin{equation}
W^t=\tilde{W}^t+K_\tau ^tW^\tau ,
\end{equation}
where $\tilde{W}^t=W^0(\bar{p}_0,\bar{q}_0)$. The corresponding iterational
series solving this equation can be written as

\begin{equation}
W^t=\tilde{W}^t+K_{\tau _1}^t\tilde{W}^{\tau _1}+K_{\tau _2}^tK_{\tau
_1}^{\tau _2}\tilde{W}^{\tau _1}+K_{\tau _3}^tK_{\tau _2}^{\tau _3}K_{\tau
_1}^{\tau _2}\tilde{W}^{\tau _1}+...
\end{equation}
Now, to calculate the quantum-mechanical average (1) it is necessary to
calculate a linear functional of the Wigner distribution function

\begin{eqnarray}
A\left( t\right) &=&\int \!\!\!\int dqdpA\left( q,p\right) W\left(
q,p,t\right)  \nonumber \\
&=&\left( A|\tilde{W}^t\right) +\left( A|K_{\tau _1}^t\tilde{W}^{\tau
_1}\right) +\left( A|K_{\tau _2}^tK_{\tau _1}^{\tau _2}\tilde{W}^{\tau
_1}\right) +\left( A|K_{\tau _3}^tK_{\tau _2}^{\tau _3}K_{\tau _1}^{\tau _2}%
\tilde{W}^{\tau _1}\right) +...
\end{eqnarray}
Here the brackets $(...\mid ...)$ for the functions $A=A(p,q)$ and $\tilde{W}%
^t$ or $K_{\tau _i}^tK_{\tau _{i-1}}^{\tau _i}...K_{\tau _1}^{\tau _2}\tilde{%
W}^{\tau _1}$ indicate averaging over the entire phase space $\{p,q\}$.

The first term on the right-hand side of Eq. (10) gives the classically
evolving initial distribution $W^0(\bar{p}_0,\bar{q}_0)$, i.e. the evolution
of the distribution function without quantum corrections. However, even this
first term of the iterational series describes not classical but rather
quantum effects and can contain arbitrary powers of the Planck constant,
since a quantum initial state of the system is taken as the initial data for
Eq. (10). The rest of the terms in the iterational series describe quantum
corrections to evolution. Each term of the iterational series (10) is a
multiple integral. This multiple integral can be replaced by an integral
sum, and each term of the integral sum can be represented as a contribution
of trajectories of a definite topological type. These trajectories consist
of segments of classical trajectories --- solutions of Eqs. (7) ---
separated from one another by random perturbations of the momentum.

All terms of the iterational series can be calculated in accordance with the
theory of Monte Carlo methods for solving linear integral equations. For
this purpose the Monte Carlo scheme was developed, which provides
the essential sampling of the iteration series terms (10). This essential
sampling also decreases the computer time required to calculate the rest of the
integrals appearing in each term of the iterational series. Let us consider
the second term of the series (10). This term can be rewritten as

\begin{eqnarray}
K_{\tau _1}^t\tilde{W}^{\tau _1} &=&\int\limits_0^1d\tau _1\int ds_1\omega
\left( s_1,\bar{q}_1\right) W^0(\bar{p}_0^1,\bar{q}_0^1)=  \nonumber \\
&=&\int\limits_0^1d\tau _1\left[ B(\bar{q}_2)\left( 1+Q(\bar{q}_2)\right)
\right] \theta \left( 1-\tau _2\right) r\left( \tau _2\right) \int ds_1P(s_1,%
\bar{q}  \nonumber \\
&\times &\left\{ \sigma \left( s_1,\bar{q}_1^2\right) t\tilde{Q}(\bar{q}%
_1^2)\theta \left( \tau _2-\tau _1\right) /C(\bar{q}_1^2)r(\tau _1)\right\}
W^0(\bar{p}_0^1,\bar{q}_0^1),
\end{eqnarray}
where the substitution of variables $\tau \rightarrow \tau t$ was made for
all terms of the iterational series (10). The quantity $r(\tau _1)$ is the
probability of choosing a random time $\tau _1$ and $\theta $ is the theta
function.

Once the second term of the series (10) is written in the form (12), it can
be given the following probabilistic interpretation. We will take advantage
of the time-reversibility of the equations of classical dynamics (7) and start the
construction of a trajectory at time $\tau =0$. At time $\tau _1$ for a
trajectory representing an arbitrary term in the iterational series a
perturbation of the momentum of the trajectory by an amount $s_1$ can occur
with probability $C(\bar{q}_1^2)$, and the probability of rejecting a
momentum perturbation is $B(\bar{q}_2)$ $(C(\bar{q}_1^2)+B(\bar{q}_2)=1)$ .
The probability $B$ for rejecting momentum jumps was introduced to make the
algorithm more flexible, so that depending on the degree of quantization of
the system a transition from quantum to classical trajectories would occur
automatically.

Since we are considering a trajectory representing the second term in the
iterational series, a perturbation of the momentum at the time $\tau _1$ was
accepted. Now it is necessary to choose in the time interval $[\tau _1,1]$ a
random value $\tau _2$ which is the time of the next attempt to perturb the
momentum. After a perturbation of the momentum by an amount $s$ we must
continue the generation of the trajectory  up to the time $\tau _2$  in
accordance with Hamilton's equations. At this time an attempt to perturb the
momentum for the second term of the iterational series must be rejected, and
we continue the generation of the trajectory up to the time $\tau =1$. The
rejected attempt of the perturbation of the momentum must be taken into account by
multiplying the weighting function of the trajectory by a compensating
factor, which stands in the braces on the right-hand side of the expression
(12). The product of the Weyl symbol of the operator under consideration and
the weighting function at different points along the trajectory gives the
time dependence of the computed quantities. Averaging over a large ensemble
of trajectories of this type gives the contribution of the second term of
the iterational series.

Similar expressions but with a large number of intermediate times on
classical trajectories when a perturbation of the momentum occurs can also
be written for the other terms in the series (10). The number of the term in
the iterational series (10), described by the given trajectory, determines
the number of momentum perturbations along the trajectory.

The final expression used to calculate the linear functional (11) is
\begin{eqnarray}
\label{b34}
A\left( t\right)&=&M\left\{ \alpha \left( A;T_{i}\right) \right\}=
\sum_{p,q}\left( \triangle p\triangle q\right) \sum_{i=0}^\infty
\sum_{j=0}^i\sum_{\tau_j}\sum_{s_j}\alpha \left( A;T_{i}\right)
\times P\left(T_{i}\right) ;\nonumber\\
\alpha \left( A;T_{i}\right)&=&A\left(p,q\right)W^0(\bar{p}_0^1,\bar{q}%
_0^1)\Omega \left(T_{i}\right),
\end{eqnarray}
where the functions $P$ and $\Omega $ are, respectively, the probability of
generating a quantum trajectory $T_i$ and the weighting function of this
trajectory.

\section{Measured quantities}

The study of the evolution of a wave packet can be taken as the starting
point for studying the temporal aspects of tunneling. The probability of
observing a wave packet or particle at an arbitrary point $X$ is determined
by the squared modulus $\left| \psi (X,t)\right| ^2\,$ of the wave function.
In a nonstationary problem this probability depends on the time and
determines the characteristic times of the wave-packet dynamics. If an ideal
detector (i.e. measurement by the detector does not disturb the wave
function), sensitive to the presence of particles, is used in the
experiment, then the average presence time measured by the detector at the
point $X$ is

\begin{equation}
\tilde{t}_X=\frac{\displaystyle \int\limits_0^\infty dt\,t\left| \psi \left(
X,t\right) \right| ^2}{\displaystyle \int\limits_0^\infty dt\left| \psi
\left( X,t\right) \right| ^2}.
\end{equation}
A description of these times can be found in Refs. \onlinecite{r31}--\onlinecite{r32}. The distribution
of presence times at the point $X$ is

\begin{equation}
\tilde{P}\left( t_X\right) =\frac{\left| \psi \left( X,t\right) \right| ^2}{%
\displaystyle \int\limits_0^\infty dt\left| \psi \left( X,t\right) \right| ^2%
}.
\end{equation}
To find the squared wave function $\left| \psi (X,t)\right| ^2\,$ it is
sufficient to calculate a quantum-mechanical average of an operator

\begin{eqnarray*}
\langle \psi \left( t\right) \left| \delta \left( \hat{q}-X\right) \right|
\psi \left( t\right) \rangle =\int dq\delta \left( q-X\right) \left| \psi
\left( q,t\right) \right| ^2=\left| \psi \left( X,t\right) \right| ^2.
\end{eqnarray*}
In the Wigner representation this is equivalent to calculation
of the integral

\begin{equation}
\langle \psi \left( t\right) \left| \delta (\hat{q}-X)\right| \psi \left(
t\right) \rangle =\int \!\!\!\int dqdp\delta \left( q-X\right) W\left(
q,p,t\right) =\int dp\,W(X,p,t).
\end{equation}
If the point $X$ is chosen to the right of the barrier, then this integral
makes it possible to calculate the squared wave function which has tunneled
through the barrier. The distribution of the ``presence'' times can be
rewritten, in accordance with Eq. (16), as

\begin{equation}
P_X\left( t\right) =\frac{\left| \psi \left( X,t\right) \right| ^2}{%
\displaystyle \int\limits_0^\infty dt\left| \psi \left( X,t\right) \right| ^2%
}=\frac{\int dpW\left( X,p,t\right) }{\displaystyle \int\limits_0^\infty
dt\int dpW\left( X,p,t\right) }.
\end{equation}
To determine the average time when the wave packet passes through a detector
at the point $X$ it is necessary to calculate the integral

\begin{eqnarray}
\left\langle t\left( X\right) \right\rangle =\int\limits_0^\infty
dt\,tP_X\left( t\right) ,
\end{eqnarray}
and the average transition time of a packet from the point $X_i$ to the
point $X_f$ will be

\begin{equation}
\left\langle t_T\left( X_i,X_f\right) \right\rangle =\left\langle t\left(
X_f\right) \right\rangle -\left\langle t\left( X_i\right) \right\rangle .
\end{equation}
If the points $X_i$ and $X_f$ are chosen on different sides of the potential
barrier, then the expression (19) can be used to estimate the tunneling time.

The main drawback of the definition (17) is that, as a rule, detectors
sensitive to a flux density and not a probability density are used in
physical experiments. Therefore a different quantity must be considered in
order to compare theory and experiment. For this, the distribution of
arrival times of a wave packet at a prescribed point in terms of the
probability flux density was introduced:\cite{r34}

\begin{equation}
P_X\left( t\right) =\frac{\left\langle \psi \left( t\right) \left| \hat{J}%
\left( X\right) \right| \psi \left( t\right) \right\rangle }{\displaystyle %
\int\limits_0^\infty dt\langle \psi \left( t\right) \left| \hat{J}\left(
X\right) \right| \psi \left( t\right) \rangle },
\end{equation}
where

\begin{equation}
\hat{J}\left( X\right) =\frac 12\left[ \hat{p}\delta \left( \hat{q}-X\right)
+\delta \left( \hat{q}-X\right) \hat{p}\right] .
\end{equation}
Of course, the definition (20) is not a real distribution function from
probability theory, since this function can assume negative values at some
points. Nonetheless the definition (20) will be a distribution function if
there is no reverse flux through the point $X$ or the flux is negligibly
small. For this the point $X$ is chosen at a sufficiently far from the
barrier. Measuring the distribution of the arrival times of a packet in
front of and beyond the barrier, the transition time through a region much
larger than the region of the potential barrier can be calculated. This time
is analogous to the asymptotic phase times\cite{r35} and, besides the
tunneling time and the packet--barrier interaction time, it also contains
the transmission time through the region where the potential barrier is
zero. These two times cannot be separated. Despite continuing discussions,
this tunneling-time problem has still not been finally solved.\cite
{r19,r20,r21,r22,r23,r24,r36,r37}

Another problem concerns the physical implementation of an experiment in
which simultaneous detection of a packet in front of and beyond a barrier
would not lead to substantial reduction of the wave function. For this
reason, ordinarily, a different quantity --- the ``time delay'' --- is
measured in experiments.\cite{r38,r39,r40,r41,r42} A time delay arises
because of the presence of a barrier and is defined as the difference of the
average arrival times of the tunneling and free packets:

\begin{eqnarray}
\Delta \tau _{arrival}(X)=\langle t_X\rangle ^{tun}-\langle t_X\rangle
^{free}.
\end{eqnarray}
The definition (20) for calculation the average arrival times gives a
reasonable estimate of the time delays measured in an experiment.

The distribution of arrival times (20) can be rewritten in the Wigner
formulation of quantum mechanics as

\begin{equation}
P_X\left( t\right) =\frac{\displaystyle \int \!\!\!\displaystyle \int
dqdpJ_X\left( q,p\right) W\left( q,p,t\right) }{\displaystyle %
\int\limits_0^\infty dt\displaystyle \int \!\!\!\displaystyle \int
dqdpJ_X\left( q,p\right) W\left( q,p,t\right) },
\end{equation}
where the Weyl symbol of the current operator $\hat{J}(X)$ is

\begin{eqnarray}
J_X\left( q,p\right) =\frac \hbar 2\sin \left( \frac{2p\left( X-q\right) }%
\hbar \right) \frac \partial {\partial q}\delta \left( q-X\right) .
\end{eqnarray}
Substituting into Eq. (20) the expression (24) and calculating the integral
over the variable $q$ by parts we obtain the expression

\begin{equation}
P_X\left( t\right) =\frac{\displaystyle \int dppW(X,p,t)}{\displaystyle %
\int\limits_0^\infty dt\displaystyle \int dppW(X,p,t)}.
\end{equation}
Comparing the expressions (17) and (25), it is easy to see that they differ
by the fact that the momentum $p$ appears in the numerator and denominator
in Eq. (25). This momentum appeared in the last expression because the
probability flux density is measured there.

\section{Simulation results}

We shall examine a series of experiments on the tunneling of an electron
with the wave function

\begin{equation}
\psi (x,0)=\frac 1{\left( 2\pi \sigma _x\right) ^{1/4}}\exp \left[ -\left(
\frac{x-x_0}{2\sigma _x}\right) ^2+ik_0x\right]
\end{equation}
through a Gaussian potential barrier

\[
V\left( x\right) =V_0\exp \left[ -\frac{(x-d)^2}{\sigma ^2}\right] .
\]
The Wigner distribution function (3) corresponding to the initial wave
function of the electron can be written as

\begin{equation}
W(p,q,0)=2\exp \left[ -\frac{(q-x_0)^2}{2\sigma ^2}\right] \exp \left[ -%
\frac{2\sigma ^2(p-\hbar k_0)^2}{\hbar ^2}\right] .
\end{equation}
The center $x_0=\left\langle \psi (x,0)\right| \hat{x}\left| \psi
(x,0)\right\rangle $ of the wave packet at zero time was chosen far enough
from the left-hand boundary of the barrier so that the probability density
beyond the barrier would be negligibly small compared with the transmission
probability $\left| T\right| ^2$ through the barrier. Tunneling occurred
through a ``wide'' $(\sigma =2.5$ nm --- this parameter of the barrier is
characteristic for Al$_x$Ga$_{1-x}$As structures) and a ``narrow'' $(\sigma
=0.5$ nm) Gaussian barrier of height $V_0=0.3$ eV centered at $d=0$. The
electron kinetic energy was $E_0=\hbar ^2k_0^2/2m=V_0/2=0.15$ eV. We used
the system of units where $\hbar =m=V_0=1$. Distances were measured in units
of the reduced de Broglie wavelength $\lambda =1/k_0$. In this system of
units the parameters of the wave packet and barrier are: $E_0=0.5,$ $\Delta
k=0.04$ $(0.125),$ $\sigma _x=1/2\Delta k=12.5$ $(4),$ $x_0=-92.5$ $(-43),$ $%
\sigma =5$ $(2.5$ nm$),$ and $\sigma =1$ $(0.5\,$nm$)$.

\subsection{Evolution of the wave packet}

The interaction of a wave packet $(\hbar \Delta k=0.125)$ with a narrow
potential barrier $(\sigma =1$ $(0.5$ nm$))\,$ is shown in Figs. \ref{f1}a
and b. These figures show the probability density $\left| \psi (x,t)\right|
^2$ (curves 1--5) of reflected (Fig. \ref{f1}a) and tunneled (Fig. \ref{f1}%
b) wave packets at successive times $t=114-239$ fs. The probability density
was calculated using Eq. (16), i.e. in terms of the Wigner distribution
function. This integral was calculated along quantum and classical
trajectories. In the calculation over classical trajectories only the
high-energy components of a packet could pass classically above the barrier.
This calculation corresponds to the curve 1 in Fig. \ref{f1}c, and the
evolution of the Wigner function can be described only by the first term of
the series (10). In the formalism of quantum trajectories the passage of the
components of a packet beyond the barrier is associated with random
perturbations of the momentum, i.e. with a virtual change in energy. The
results of this calculation correspond to the curve 2 in Fig. \ref{f1}c. Now
the quantum corrections introduced by all terms in the series (10) are taken
into account in the evolution of the Wigner function.

Of course, the calculation over quantum trajectories also takes account of
the high-energy components that pass above the barrier, since they
describe the contribution of the first term in the series (10). However,
comparing the curves 1 and 2 in Fig. \ref{f1}c shows that their role is
negligible for a narrow barrier and most of the packet passes above the
barrier on account of the virtual change in energy, described as random
perturbations of the momentum of the quantum trajectories. A study of
tunneling through a wide barrier leads to the opposite conclusion. The
curves 1 and 2 in Fig. \ref{f1}d approximately coincide. This means that most
of the packet has passed above the barrier, and the contribution of all
terms in the series (10), except for the first term, is negligibly small. To
avoid such a situation and to restore the importance of quantum effects, it
is necessary to decrease the uncertainty of the momentum of the initial wave
packet. In what follows all calculations for a wide barrier are presented
for momentum uncertainty $\hbar \Delta k=0.04$.

\subsection{Average coordinate, average momentum, and their variances}

Figure \ref{f2}a shows the evolution of the average coordinate $\left\langle
\psi (t)\right| \hat{X}\left| \psi (t)\right\rangle $ of the wave packet for
calculation according to classical (curve 1) and quantum (curve 2)
trajectories. In these two methods for calculation the average coordinate $%
\bar{X}$ no differences are observed before interaction with the barrier
(curves 1 and 2 are coincident). This result can be explained as follows. In
the method under discussion the quantum-mechanical properties appear at two
points: in the properties of the initial state of a wave packet and in the
evolution of the packet. Since the same initial data were chosen for the
quantum and classical trajectories, the fact that $\bar{X}$ is the same must
be explained by the evolution of the wave packet. Specifically, while the
packet moves freely in front of the barrier, it is correctly described by
classical trajectories also. In this case the first term in the series (10)
is sufficient to describe the evolution of the Wigner function. This result
can also be obtained analytically, estimating the right-hand side of the
Wigner--Liouville equation (4). For the initial Wigner function (27) and
Gaussian barrier which we have chosen it is easy to show that the integral
on the right-hand side of Eq. (4) decays exponentially with increasing
distance from the barrier. In this case Eq. (4) becomes the classical
Liouville equation, whose characteristics are ordinary classical
trajectories.

A difference in the behavior of the curves 1 and 2 appears after the packet
interacts with a barrier. Now the classical trajectories are no longer
characteristics and do not describe the evolution of the wave packet
correctly. In Figs. \ref{f2}a and b the average coordinate and the momentum
of the calculation over quantum trajectories (curve 2) are greater than for
classical trajectories (curve 1). This is due to the following
circumstances. In the first place, since most of the packet is reflected, as
one can see from Fig. \ref{f2}b the average momentum changes sign after
being scattered by the barrier. In the second place, the classical
trajectories (curve 1) do not take account of tunneling; they only take
account of the negligible above-barrier transmission, arising because of the
uncertainty in the momentum of a Gaussian wave packet. At the same time it
is obvious that the tunneling part of the packet has positive momentum and
moves in the opposite direction relative to the reflected part. Therefore
its contribution to $\bar{X}$ and $\bar{P}\,$ has a different sign. This is
the explanation of the difference between the curves 1 and 2.

In addition, the motion of the tunneling and reflected packets on different
sides of the barrier also explains the more rapid increase of the coordinate
variance in the quantum calculation (curve 2, Fig. \ref{f2}c) as compared
with the classical calculation (curve 1), which takes into account only the
spreading of the wave packet. The behavior of the packet width on scattering
by a barrier is shown in greater detail in the upper left-hand part of Fig.
\ref{f2}c.

The interaction of a packet with the barrier also leads to an interesting
behavior of the momentum variance in Fig. \ref{f2}d. The constant values
(curve 1) on the initial and final sections show the momentum variance in
the incident and reflected wave packets, i.e. before and after interaction
with the barrier. The observed peak is due to the change in the sign of the
momentum of the packet and to the fact that different components reach the
barrier and are reflected from it at different times. The increase in the
momentum variance (curve 2) on the final section is explained by the
appearance of a tunneling packet with positive momentum in the quantum
computational method, while the total average momentum is negative.

\subsection{Distribution of arrival and presence times. Momentum
distribution function}

The results of the calculation of the unnormalized presence time distribution
(17) at different points in front of the barrier, inside the barrier,
and beyond barrier are presented in Figs. \ref{f3}a and b (curves 1--5).
Figures \ref{f4}a and b show the analogous results for the unnormalized
distribution (20) of the arrival times. The curves 1 in Figs. \ref{f3}a and
\ref{f4}a show the behavior of the probability density and flux,
corresponding to the fact that the incident and reflected wave packets pass
through the detector at different times. Curve 2 in Fig. \ref{f4}a shows the
behavior of the flux measured at a certain point to the left of barrier
center. The tunneling and high-energy components present in the initial
packet can classically reach this point. An
interesting result is obtained for the probability flux density in Fig. \ref
{f4}b (curves 3--5). The flux measured at barrier center (curve 3) is much
less than the flux on the right-hand boundary of the barrier (curve 4) and
far to the right of the barrier (curve 5). This means that interference of
the tunneling components of the wave packet which move in opppostie
directions occurs inside the barrier. Some of these components pass
completely through the barrier, while others are reflected inside the
barrier and do not reach its right-hand boundary. Interference of the
reflected and transmitted components leads to the observed decrease in the
flux amplitude at barrier center (curve 3) and at the right-hand boundary
(compare curves 4 and 5). It is interesting that the investigation of tunneling
using classical trajectories in complex time also reveals the similar effect.%
\cite{r37} It was found that transition through a barrier occurs as a
series of attempts, many of which are unsuccessful because of reflections in
different regions under the barrier.

The comparison of the presence and arrival times distributions in Figs. \ref
{f3}b and \ref{f4}b shows that they are almost identical. The computed
average presence and arrival times (18) are also identical (the difference
is less than 1 fs). As we have already stated, the distribution of the
arrival times (20) is not a true distribution function and, as one can see
from Fig. \ref{f4}a (curve 2), it is not suitable for calculation of the
average arrival time of a packet in front of the barrier. This makes it
impossible to calculate the tunneling time as the difference (19) of the
average arrival times of the packet in front of and beyond the barrier.
Nonetheless the expression (19) can be used to estimate the tunneling time,
if the average presence time (14) is used instead of the average arrival
time in front of the barrier. Then the tunneling time through the potential
barrier is $\tau _T(-0.67\sigma ,+0.67\sigma )=12$ fs, i.e. it is almost
equal to the transmission time of a free packet through a similar region $%
\tau _T^{class}(-0.67\sigma ,+0.67\sigma )=13.5$ fs.

The time delays were measured at the points $x_4=0.67\sigma $ $(1.6$ nm) and
$x_5=5\sigma $ (12 nm) and were found to be $\Delta \tau _{arrival}(x_4)=8$
fs and $\Delta \tau _{arrival}(x_5)\le 0.5$ fs. If these measurements were
performed even farther to the right of the barrier, then $\Delta \tau
_{arrival}(x)$ would become negative. Thus an interesting behavior is discerned:
Even though the tunneling wave packet is delayed by the barrier ($\Delta
\tau _{arrival}(x_4)=8$ fs) and passes through the barrier approximately in
the same time as a free packet, it appears earlier at a definite distance to
the right of the barrier. This effect can be explained by the fact that the
transmission probability through a Gaussian barrier increases with energy,
so that packet components with a larger momentum have a higher probability
of ending up beyond the barrier. These components move more rapidly than a
free packet and eventually overtake a free packet. Then the time delays can
only be negative. This confirms the momentum distribution function

\begin{equation}
\frac{\langle \psi \left( t\right) \left| \delta \left( \hat{p}-p\right)
\right| \psi \left( t\right) \rangle }{\langle \psi \left( t\right) |\psi
\left( t\right) \rangle },
\end{equation}
calculated for narrow (Fig. \ref{f5}a) and wide (Fig. \ref{f5}b) barriers,
respectively, at times $t=218$ and $385$ fs. At these characteristic times
the distribution function no longer changes, since the interaction with the
barrier has finished. It is evident from Fig. \ref{f5} that the average
momentum of the tunneled wave packet (curve 2) is greater than the average
momentum of the wave packet initially (curve 1). The peak observed in the
momentum distribution function (curve 2 in Fig. \ref{f5}a) is due to the
packet components that had a quite large momentum and passed above the
barrier. It is evident that tunneling through a narrow potential barrier
leads to a larger variance of the distribution function, while tunneling
through a wide barrier substantially shifts the center of the distribution
in the direction of large momenta (curve 2 in Fig. \ref{f5}b).

\section{Conclusions}

The quantum generalization of classical molecular dynamics was used to solve
the Wigner--Liouville integral equation in the Wigner formulation of quantum
mechanics. The method discussed for solving this equation does not require a
large increase of computer time and makes it possible to avoid the
computational difficulties that arise when solving the nonstationary
Schr\"{o}dinger equation.

This approach was used to solve the nonstationary problem of tunneling of a
finite wave packet, i.e. a problem in which it is important to take account
of exponentially small quantum effects. The evolution of a wave packet, the
behavior of the average and the variances of the coordinate and momentum and
the distributions of the presence and arrival times for the wave packet at
different positions of an ideal detector were analyzed. The following
results were obtained: 1) The tunneling time through a potential barrier is
approximately of the same order of magnitude as the transmission time of a
free wave packet over a similar distance; 2) the tunneling wave packet is
delayed by the potential barrier, so that after the barrier the time delay
should be positive; 3) measurement of negative time delays is possible only
at sufficiently large distances from the barrier and is associated with a
shift of the momentum distribution function; 4) a Gaussian barrier transmits
mainly the high-energy components of a packet, interaction with the
barrier shifts the center of the momentum distribution function so that the
average momentum of the transmitted packet is larger than the initial
average momentum of the entire packet; 5) tunneling through a narrow
potential barrier leads to a larger variance of the momenta of the tunneled
components, while tunneling through a wide barrier leads to an appreciable
increase in the average momentum; and, 6) the computational results for the
probability flux density showed that the tunneling wave packet does not pass
completely through the barrier, a portion of the packet under the barrier
is reflected and does not reach its right boundary.

This work was partically supported by grants from the Russian Fund for
Fundamental Research and the Program ``Physics of Solid-State
Nanostructures.''

\newpage
\begin{figure}
\caption[]{Probability density $\left| \psi _{ref}(x,t)\right| ^2$ (a)
of the reflected wave packet (a) and probability density $\left| \psi
_{tr}(x,t)\right| ^2$ (b) of the tunneled wave packet (b) at successive
times $t_i=144-239$ fs (curves 1--5) with $\Delta k=0.125$ and barrier
``thickness'' $\sigma =1$ (0.5 nm); $\left| \psi _{tr}(x,t)\right| ^2$ (c,
d) at time $t=187$ fs, $\Delta k=0.125$ with barrier thickness $\sigma =1$
(0.5 nm) (c) and $\sigma =5$ (2.5 nm) (d): curve 1 --- calculation using
classical trajectories, curve 2 --- calculation using quantum trajectories.}%
\label{f1}
\end{figure}

\begin{figure}
\caption[]{Average coordinate $\bar{X}$ (a), average momentum $\bar{P}$
(b), coordinate variance\thinspace  $\left\langle (X-\bar{X})^2\right\rangle
$ (c) and momentum variance\thinspace  $\left\langle (P-\bar{P}%
)^2\right\rangle $ (d): 1 --- calculation using classical trajectories; 2
--- calculation using quantum trajectories.}
\label{f2}
\end{figure}

\begin{figure}
\caption[]{Probability density or unnormalized distributions of the
presence times (17): a --- $\left| \psi (x_i,t)\right| ^2$ at the point $%
x_1=-5\sigma $ (curve 1); b --- $\left| \psi (x_i,t)\right| ^2$ at the point
$x_2=-0.67\sigma $ (curve 2), at $x_4=0.67\sigma $ (curve 4), and $%
x_5=5\sigma $ (curve 5); the center of the barrier is located at $x_3=0$,
the barrier ``thickness'' $\sigma =5$ (2.5 nm).}
\label{f3}
\end{figure}

\begin{figure}
\caption[]{Flux of probability density or unnormalized distributions of the
arrival times (20): a --- $J(x_i,t)$ at the points $x_1=-5\sigma $ (curve 1)
and point $x_2=-0.67\sigma $ (curve 2); b --- $J(x_i,t)$ at $x_3=0$ (curve
3), $x_4=0.67\sigma $ (curve 4), and $x_5=5\sigma $ (curve 5); the center of
the barrier is located at $x_3=0$, the barrier ``thickness'' $\sigma =5$
(2.5 nm).}
\label{f4}
\end{figure}

\begin{figure}
\caption[]{Momentum distribution in a packet at $t=0$ (curve 1) and in a
packet transmitted through a potential barrier (curve 2): a --- $\Delta
k=0.125,$ barrier ``thickness'' $\sigma =1$ (0.5 nm), $t=218$ fs; b --- $%
\Delta k=0.04,$ barrier ``thickness'' $\sigma =5$ (2.5 nm), $t=385$ fs.}
\label{f5}
\end{figure}

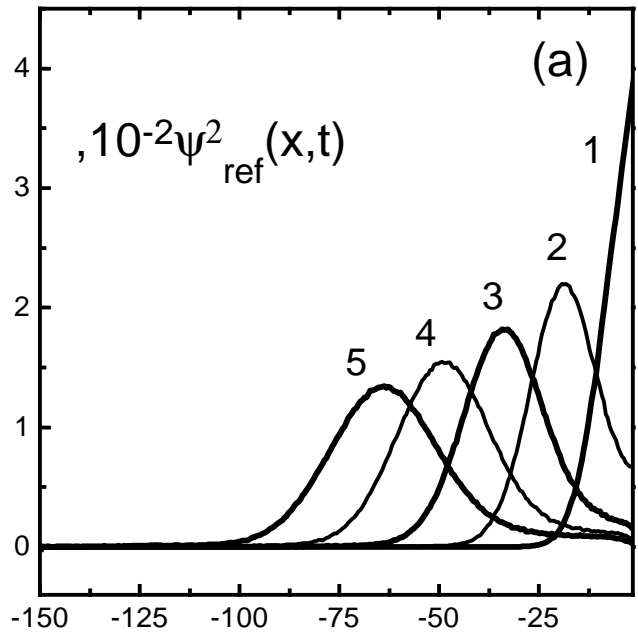
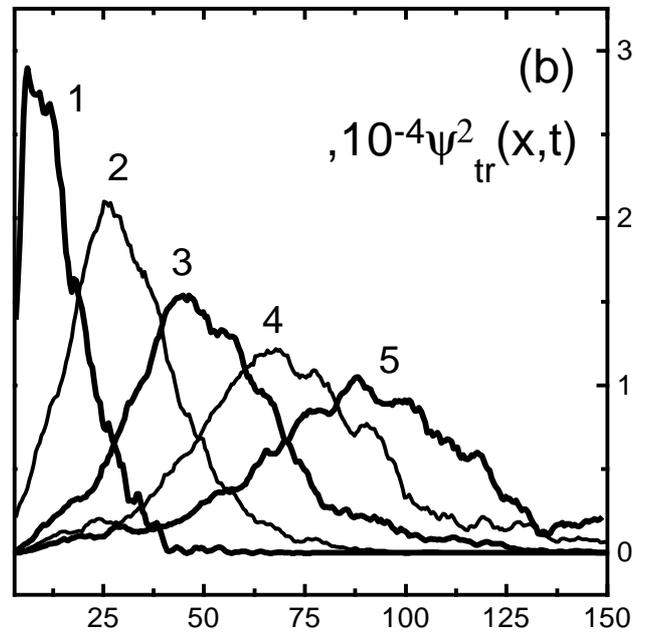
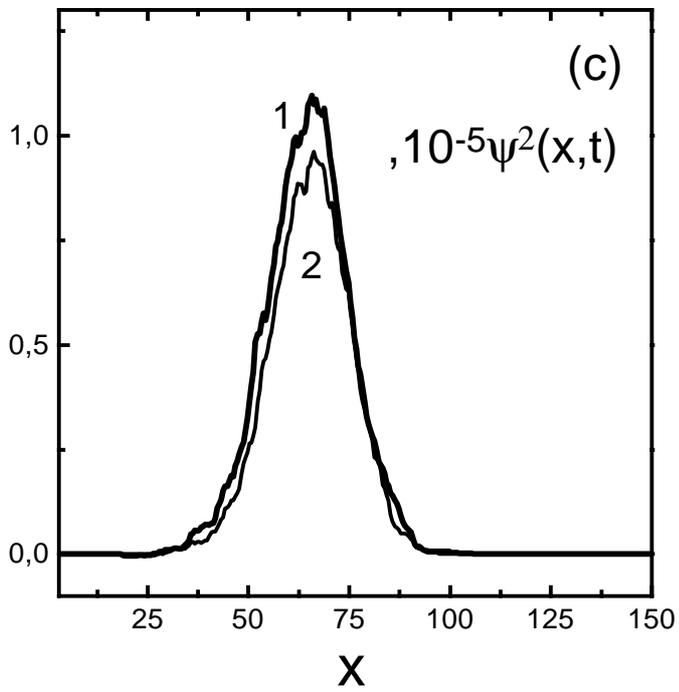
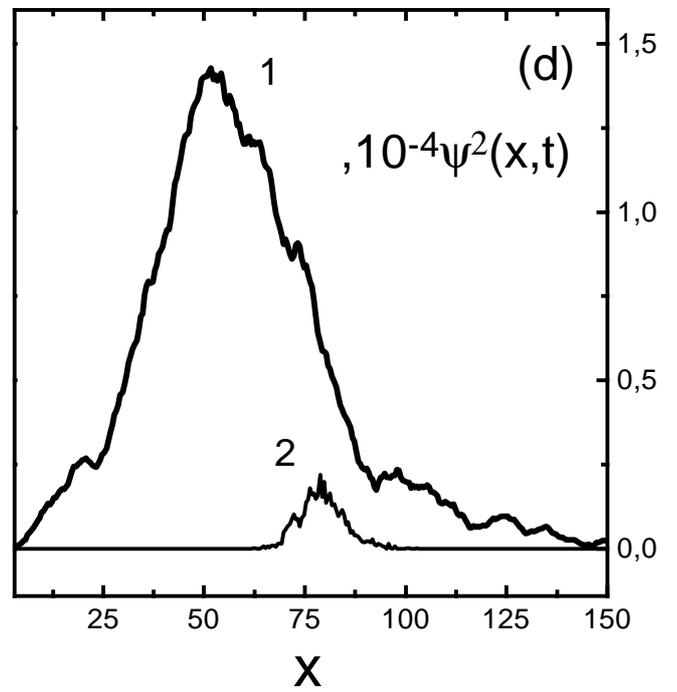
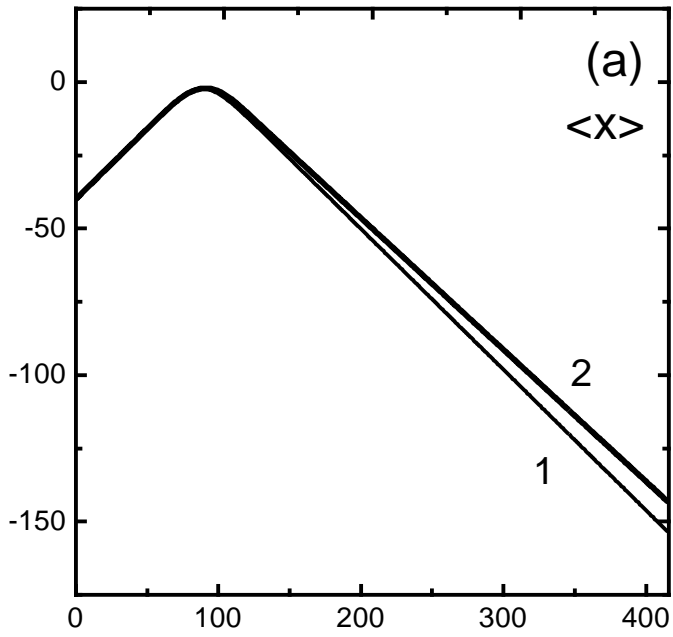
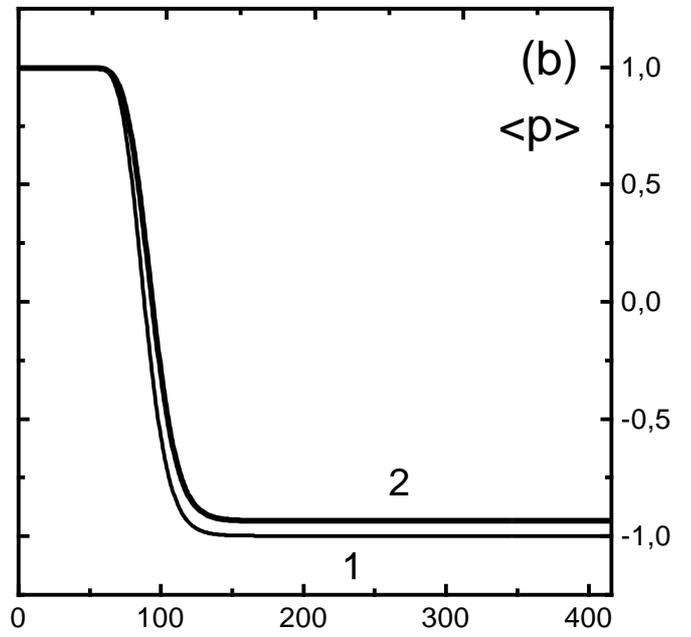
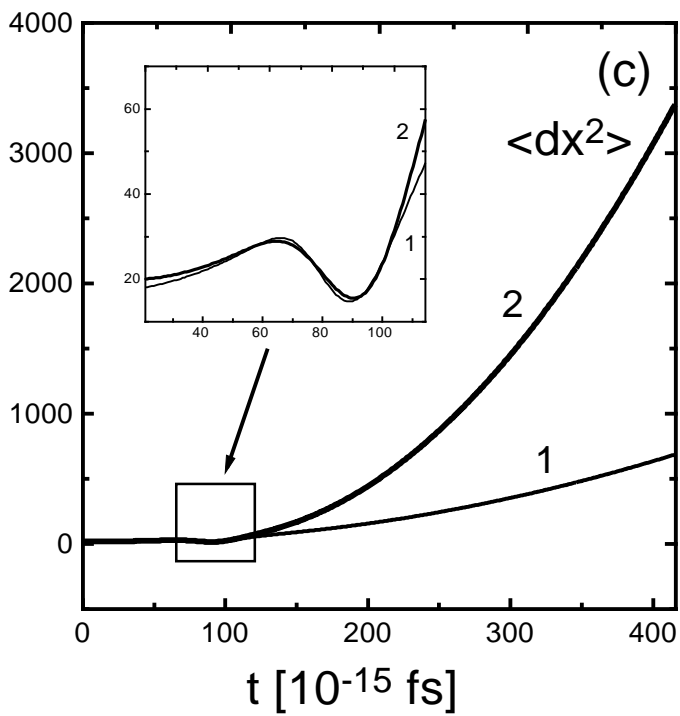
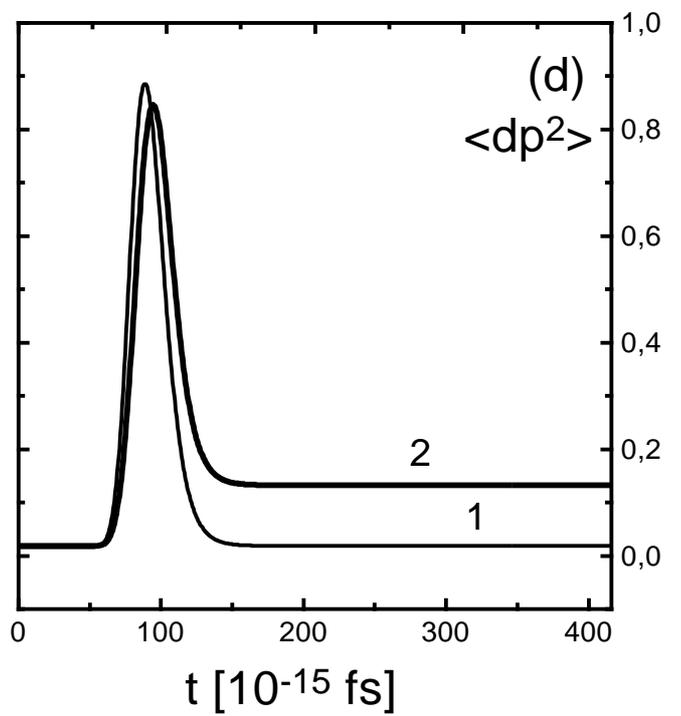
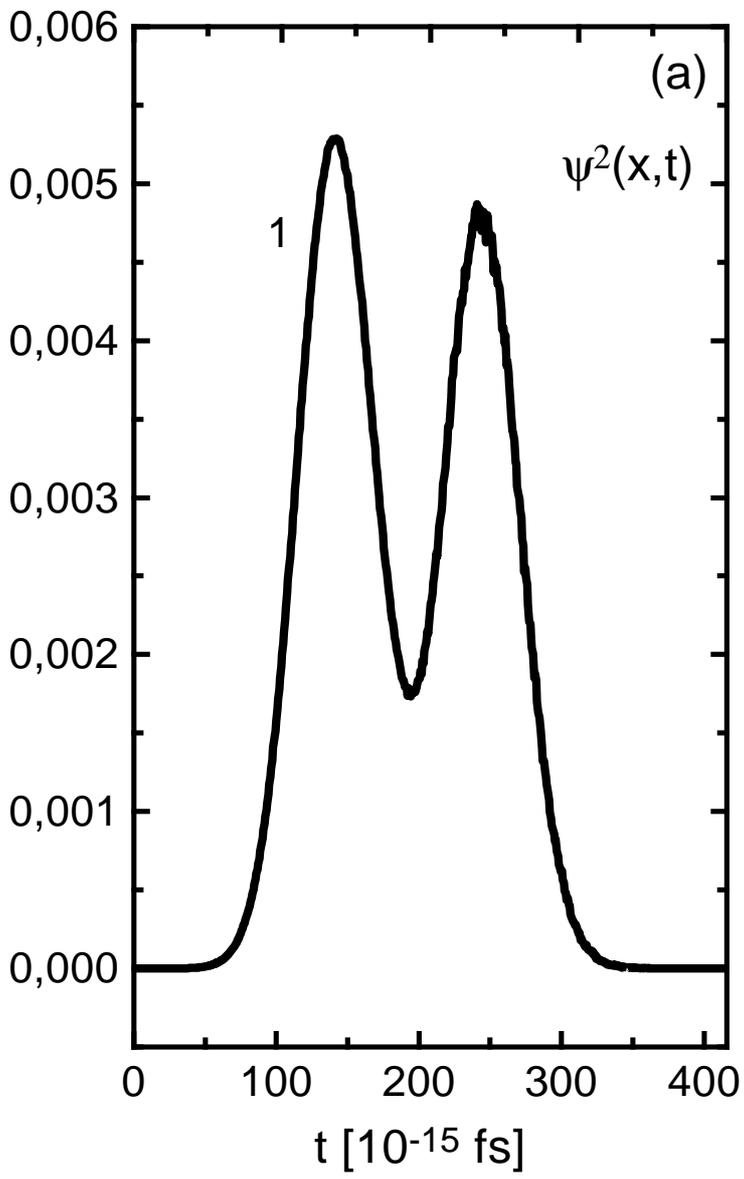
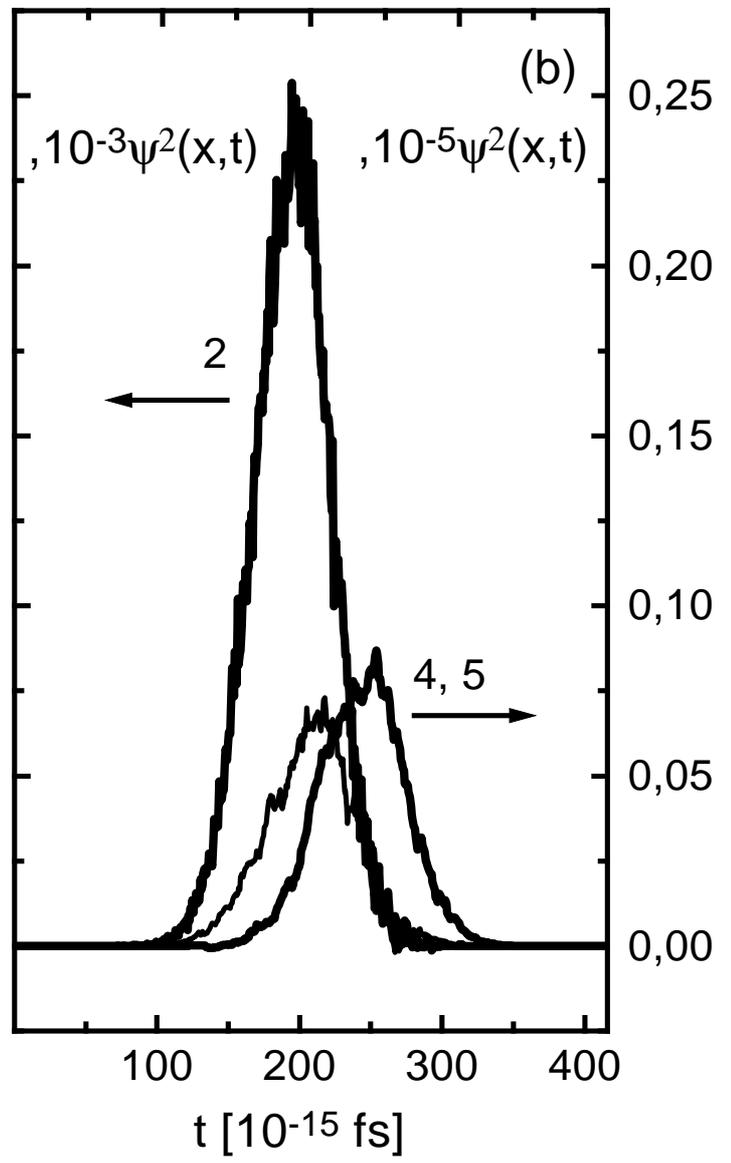
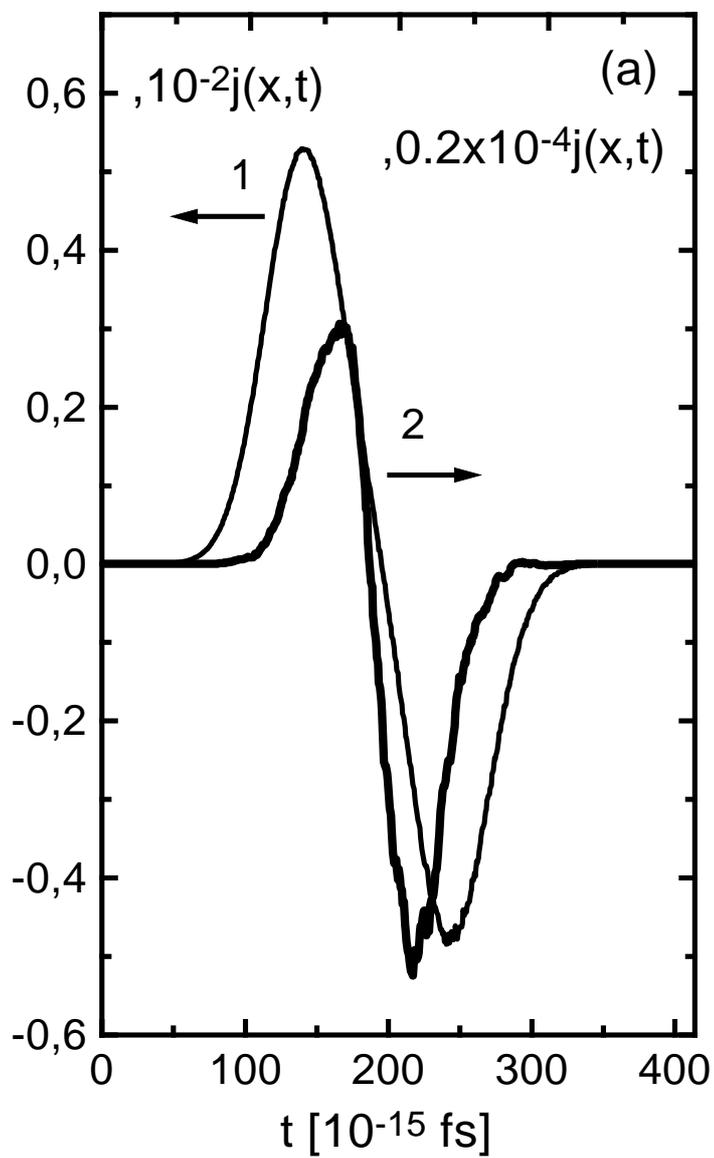
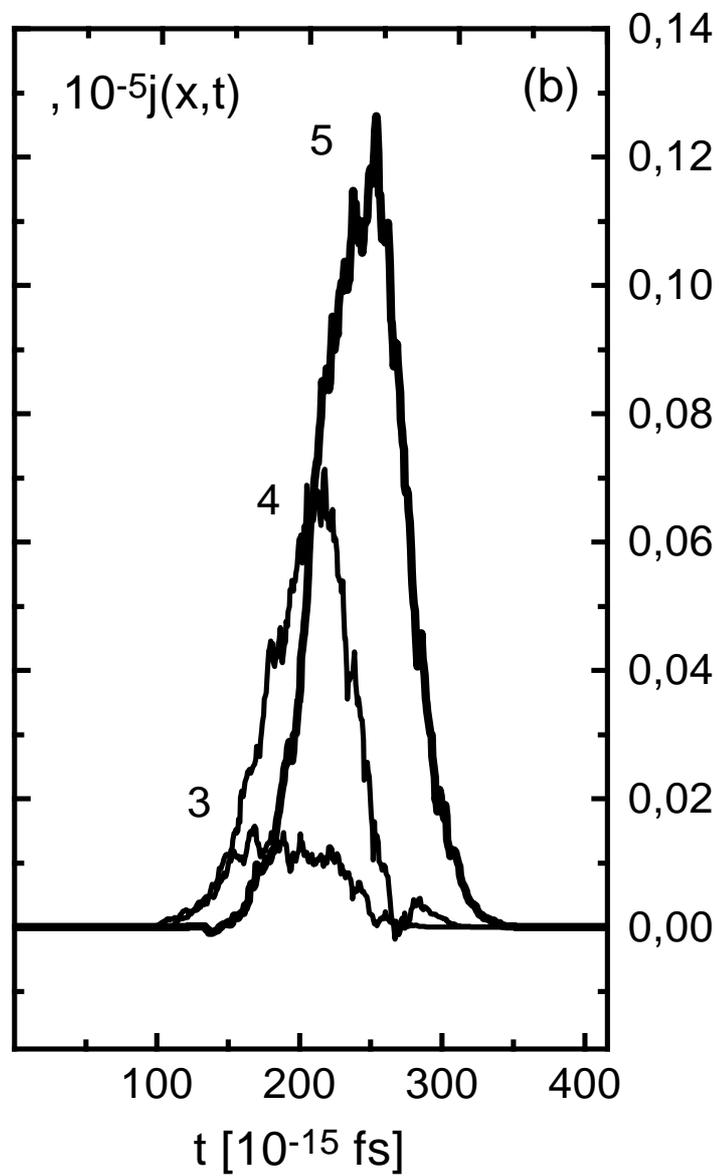
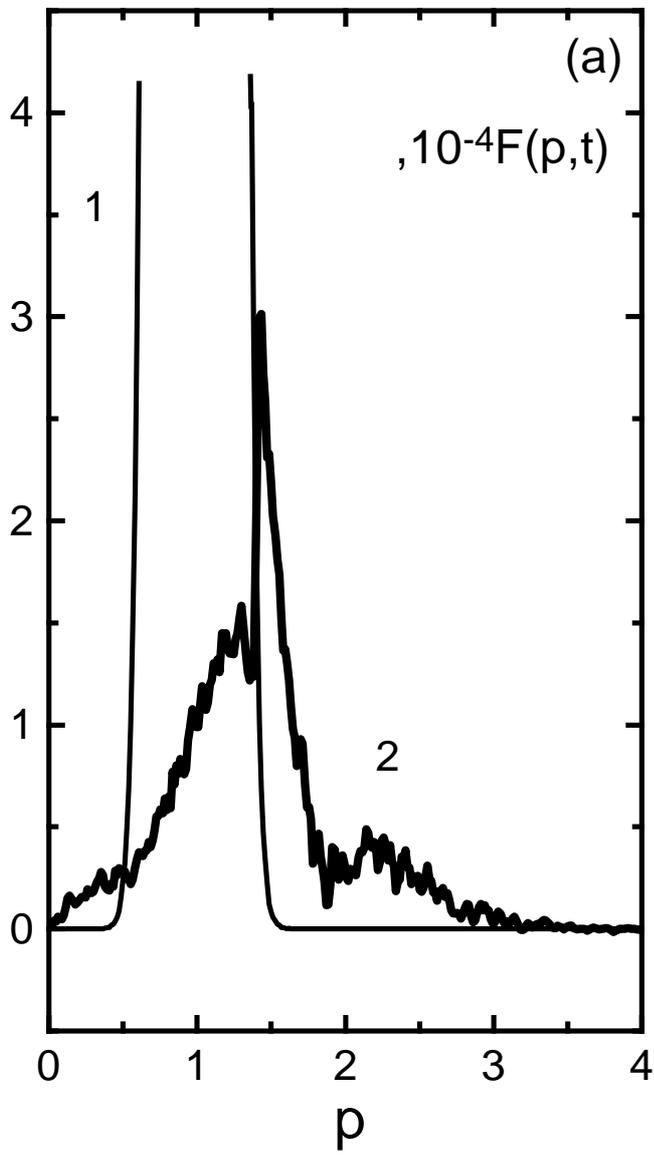
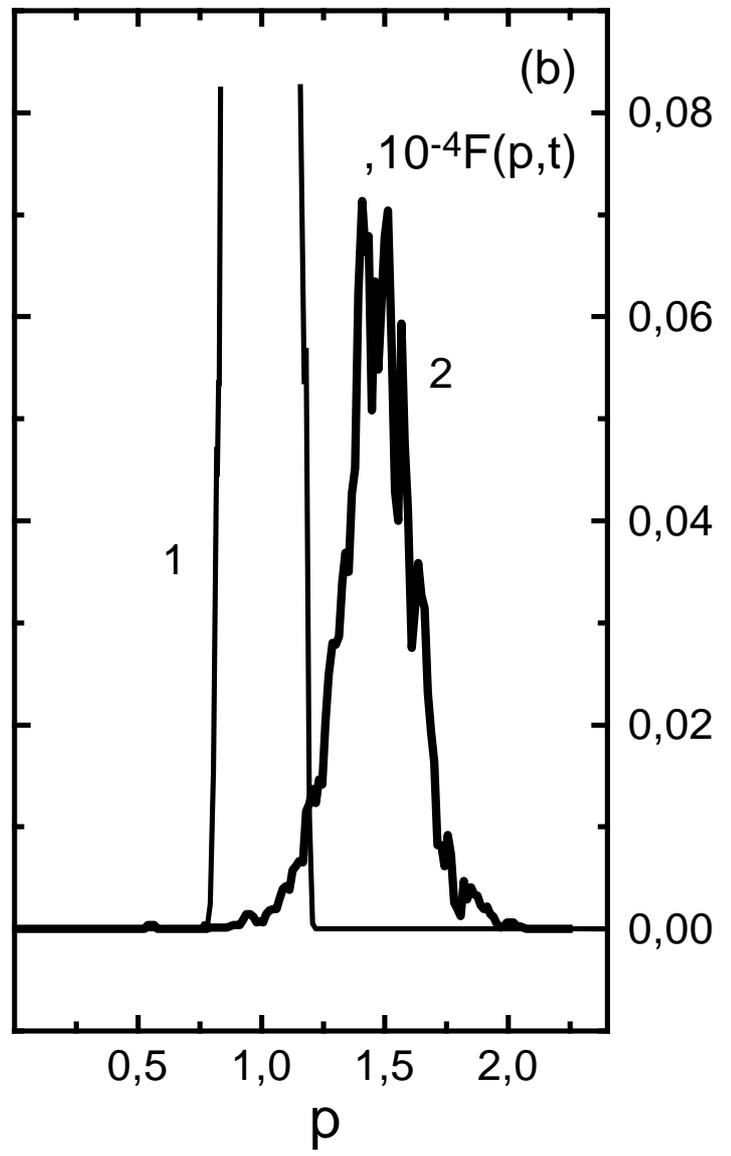
\end{document}